\documentclass[aps,prl,twocolumn,superscriptaddress,showpacs]{revtex4}

\usepackage{amssymb}
\usepackage{amsmath}

\newcommand{\beq}{\begin{equation}}
\newcommand{\eeq}{\end{equation}}
\newcommand{\nec}{\newcommand}
\nec{\cts}{conformal transformations }
\nec{\fourg}{g_{\alpha\beta}} 
\nec{\grad}{\bigtriangledown}
\nec{\fourr}{^{(4)}R}
\nec{\detg}{^{(4)}g}
\nec{\eins}{\Biggl(R_{\alpha\beta} - \frac{1}{2}g_{\alpha\beta}R\Biggr)}
\nec{\kk}{K^{ab}K_{ab}}
\nec{\bec}{\begin{center}}
\nec{\eec}{\end{center}}
\nec{\bb}{B^{ab}B_{ab}}
\nec{\rp}{R - 8\frac{\grad^{2}\psi}{\psi}}
\nec{\pipi}{\pi^{ab}\pi_{ab}}
\nec{\beqq}{\begin{equation*}}
\nec{\eeqq}{\end{equation*}}
\nec{\V}{V(\psi)}
\nec{\delg}{\frac{\partial g_{ab}}{\partial t}}
\nec{\wt}{\widetilde}
\nec{\gt}{\longrightarrow}
\nec{\wh}{\widehat}

\begin{document}


\title{A first-principles derivation of York scaling and the 
Lichnerowicz--York equation}



\author{E. Anderson}
\email{eda@maths.qmul.ac.uk}
\affiliation{Astronomy Unit, School of Mathematical Sciences, 
Queen Mary University of London, 
London, E1 4NS, UK}

\author{J. Barbour}
\email{julian@platonia.com}
\affiliation{The Leibniz Institute at College Farm, South Newington, Banbury, Oxon, 0X15 4JG, UK}

\author{B.Z. Foster}
\email{bzf@physics.umd.edu}
\affiliation
{Institut d'Astrophysique de Paris, 98 bis Bvd. 
Arago 75014 Paris, France}
\affiliation{Physics Department, Univ. of Maryland, College Park,  
Maryland, USA}

\author{B. Kelleher}
\email{bk@physics.ucc.ie}

\author{N. \'{O} Murchadha}
\email{niall@ucc.ie}
\affiliation{Physics Department, University College Cork, Ireland}

\date{\today}

\begin{abstract}
The only efficient and robust method of generating 
consistent initial data in general relativity 
is the conformal technique initiated by Lichnerowicz 
and perfected by York. In the spatially compact 
case, the complete scheme consists of the 
Arnowitt--Deser--Misner (ADM) Hamiltonian and 
momentum constraints, the ADM Euler--Lagrange 
equations, York's constant-mean-curvature (CMC) 
condition, and a lapse-fixing equation (LFE) that 
ensures propagation of the CMC condition by the 
Euler--Lagrange equations. The Hamiltonian 
constraint is rewritten as the Lichnerowicz--York 
equation for the conformal factor $\psi$ of the 
physical metric $\psi^4g_{ij}$ given an initial 
unphysical 3-metric $g_{ij}$. The CMC condition 
and LFE introduce a distinguished foliation 
(definition of simultaneity) on spacetime, and 
separate scaling laws for the canonical momenta and 
their trace are used.  In this article, we derive 
all these features in a single package by seeking a 
gauge theory of geometrodynamics (evolving 
3-geometries) invariant under both three-dimensional 
diffeomorphisms and volume-preserving conformal 
transformations.
\end{abstract}


\maketitle


One traditionally views general relativity~(GR) as giving 4-d space-times satisfying 
the Einstein equations.  However, to make easier contact with the rest of physics, one would like 
to study GR, and gravitation more generally, as a dynamical theory.  In the standard way of doing 
this, due to Arnowitt--Deser--Misner~(ADM)~\cite{adm} and Dirac~\cite{Dirac}, 
one introduces some time-function into the space-time.  Each time-slice is a three-dimensional 
Riemannian manifold with a 3-metric $g_{ij}$.  One can view the dynamics of GR as the evolution of 
these 3-metrics.  In this letter, we will restrict our attention to smooth metrics on a fixed 
manifold that is compact without boundary~(CWB). 

The 4-d general covariance of the original theory allows free 3-d coordinate 
transformations on each slice.  This means that the configuration space is {\it superspace}, 
the space of 3-geometries, identified as the quotient with respect to 3-diffeomorphisms of 
Riem, the space of all 3-metrics defined on a given CWB manifold.  GR in this approach is 
a theory with constraints.  One of the constraints, the {\it momentum constraint} is a 3-vector 
condition that just generates the 3-diffeomorphisms.  The other constraint, the 
{\it Hamiltonian constraint}, is a scalar that represents the freedom one has in choosing 
the time-function.  A 3-metric has six free components; factoring out by the 3-diffeomorphisms 
and the choice of time-function leaves two degrees-of-freedom per space point, exactly as we 
expect for the gravitational field.  Finding an explicit representation of these two 
degrees-of-freedom is of major practical and theoretical interest.

Conformal 3-geometries have exactly two degrees-of-freedom per space point. 
They are obtained by identifying 3-geometries equivalent up to {3-d} conformal transformations, 
$g_{ij} \rightarrow\omega^4 g_{ij}$, where $\omega$ is an arbitrary suitably continuous function.  
One obtains {\it conformal superspace}~(CS)~\cite{CS}, the space of conformal 3-geometries, 
by quotienting superspace 
by the conformal transformations.  

In the York method~\cite{yor, MTW}, one constructs initial data satisfying the 
constraints in the form that they take on a constant mean 
curvature~(CMC) slice.  One picks a 3-metric, a transverse 
traceless~(TT) symmetric two-index tensor on the 3-manifold, and a single number, the value of the CMC.  
There is a distinguished rescaling of York's TT tensors under conformal transformations. Its use makes it possible to
solve the momentum constraint algebraically, and the Hamiltonian constraint reduces to a well-behaved elliptic 
scalar equation for the conformal factor $\psi$, the {\it Lichnerowicz--York equation}~\cite{Lich, yor}.  
Because this is to be solved for the conformal factor, the original 3-metric $g_{ij}$ 
turns out to be unphysical, but the physical 3-metric, $\psi^4g_{ij}$, is deduced 
by solving for the conformal factor.  One can adjoin to the above constraint system the 
{\it lapse-fixing equation}~(LFE), another well-behaved elliptic equation. This allows one to construct 
a CMC foliation of the spacetime and thus generate a 4-manifold satisfying the Einstein equations.   

Here, we derive the salient features of the York method from first principles that have not hitherto been used in gravitational theory.  We will first 
explain our principles, define our action, discuss our variational techniques, and derive our 
results.  We conclude by comparing our approach with York's.    

We wish to construct a dynamical theory of 3-geometry by extremalizing an action defined on a suitable 
configuration space.  Two general principles underlie our construction: 
1) Time is derived from change;
2) Motion and size are relative.   
Principle 1 leads us to construct on our configuration space a theory of `timeless geodesics', 
parametrized by a freely chosen label $\lambda$. Local proper time is emergent in this 
approach. Principle 2 leads us to build into the theory invariance under $\lambda$-dependent 
3-diffeomorphisms and conformal transformations.  Our principles will allow us to derive York's 
approach to GR (in the CWB case) without any prior assumptions of 4-d space-time structure. 

Consider the Baierlein--Sharp--Wheeler~(BSW)~\cite{bsw} form of the GR action:
\beq 
	I_{\mbox{\scriptsize BSW\normalsize}} 
	= \int\textrm d\lambda
		\mathcal L_{\mbox{\scriptsize BSW\normalsize}}
	=\int {\textrm d}
	\lambda{\textrm d}^3x \sqrt{g}\sqrt{R}\sqrt{T},
\label{1} 
\eeq
where $g=\textrm{det} (g_{ij})$, $R$ is the 3-d scalar curvature, and 
\beq
T = G^{ijkl}\frac{d g_{ij}}{d\lambda}\frac{d g_{kl}}{d\lambda}
    \equiv G^{ijkl}(\dot{g}_{ij} - \pounds_{\xi} g_{ij})(\dot{g}_{kl} - \pounds_{\xi}g_{kl}).
\eeq
Here 
$G^{ijkl} \equiv (g^{ik}g^{jl} - g^{ij}g^{kl})$ is the inverse DeWitt supermetric, 
and $\pounds_{\xi}g_{ij}= \nabla_{i}\xi_{j} + \nabla_j\xi_i$ is the Lie derivative of the metric with respect to 
$\xi^i$. This action is invariant under local reparametrization of $\lambda$, 
satisfying Principle 1.  The emergent proper time $t$ is related to the label 
$\lambda$ by $\delta t=N\delta\lambda$, for $N \equiv \frac{1}{2}\sqrt{T/R}$.  
The covector $\xi_i = g_{ij}\xi^j$ is effectively a gauge auxiliary that renders the action invariant 
under 3-diffeomorphisms, satisfying the `motion is relative' part of Principle 2.  

The BSW action is defined on curves in superspace.  We considered generalizations of this action 
in~\cite{rwr}.  As this led us almost uniquely back to the BSW action of GR, it constitutes a new 
derivation of GR based on our principles.  In~\cite{bom}, we extended the BSW action so as to be 
defined on CS, thus satisfying also the `size is relative' part of Principle 2.  One could 
hope that this move would lead to GR in the York picture---in fact, this is almost correct.  
While this leads to a self-consistent theory of gravity, it is not GR.  Here, we shall instead extend 
the BSW action to include invariance under conformal transformations {\it that preserve the global 
volume} of the 3-metric.  This volume-preserving condition allows us to recover standard GR in the 
York picture.  

We will denote a volume-preserving conformal transformation~(VPCT) with a `hat': 
\beq
	g_{ij}(x) \rightarrow \wh{\omega}(x)^4g_{ij}(x).
\eeq  
One can construct a
VPCT  $\wh\omega$ from any unrestricted conformal transformation $\omega$:
\beq
	\wh{\omega} = \frac{\omega}{<\omega^6>^{1/6}}, 
\label{VPCT}
\eeq
where 
\beq
	<F> = \frac{\int\textrm{d}^3x\sqrt{g}\,F}
		{\int \textrm{d}^3x\sqrt{g}},
\eeq 
denotes the global average of some function $F$.  
One can express any VPCT $\wh{\omega}$ in this manner. 

We implement the conformal symmetry in the BSW action~\eqref{1} by introducing an auxiliary scalar 
field $\phi$, constructing from it in the manner above the field $\wh\phi$.  Under a VPCT~(\ref{VPCT}), 
we declare that $\wh\phi$ transforms as
\beq
	\wh\phi \rightarrow \frac{\wh\phi}{\wh\omega}.
\label{y}
\eeq
Consequently, the `corrected coordinates' $\overline g_{ij}\equiv\wh\phi^4 g_{ij}$ are invariant 
under a VPCT.  We then re-express the BSW action in terms of the variables $\overline g_{ij}$, 
obtaining an action functional of the variables $g_{ij}$ and $\phi$.  We emphasize that we vary $\phi$ freely, not $\wh\phi$.

On the introduction of the auxiliary variable $\phi$, the constituent parts of the BSW action become
\beq
\begin{split} 
    R&\rightarrow \wh\phi^{-4}\,
	\biggl(R 
		- 8\frac{\grad^{2}\wh{\phi}}{\wh{\phi}}\biggr),
	\\T &\rightarrow \wh{T} 
	= \wh{\phi}^{-8}G^{ijkl}
		\frac{d\wh{\phi}^{4}g_{ij}}{d\lambda}
		\frac{d\wh{\phi}^{4}g_{kl}}{d\lambda}, 
\end{split}
\eeq
with
\beq 
	\frac{d\wh{\phi}^{4}g_{ij}}{d\lambda} 
	=  \wh{\phi}^{4}\left[\dot{g}_{ij} 
		- \pounds_{\xi}g_{ij} 
		+ \frac{4}{\wh{\phi}}(\dot{\wh{\phi}} 
		- \pounds_{\xi}\wh{\phi})g_{ij}\right].
\label{straight}
\eeq 
The action~\eqref{1} becomes 
\beq 
	I_{\textrm{CS+V}} 
	= \int\textrm{d}\lambda \int\textrm{d}^3x\sqrt{g}\,
	\wh\phi^{4}\,
	\sqrt{R - 8\frac{\grad^{2}\wh\phi}{\wh{\phi}}}
	\sqrt{\wh{T}}.
\label{action} 
\eeq

We now explain the rules by which we vary the action. A curve through the configuration space 
specifies a sequence $g_{ij}(\lambda)$ through Riem, combined with a $\lambda$-dependent conformal 
factor $\wh\phi(\lambda)$ and a $\lambda$-dependent coordinate transformation $\chi_i(\lambda)$ (for 
which $\xi_i = \partial \chi_i/\partial\lambda$).  We identify $\phi$ and $\chi_i$ as gauge variables, 
so they need not reduce to the identity at the end points of the curve.  Therefore, we extremize the 
action under fixed-end-point variations of $g_{ij}$ and free-end-point variations of $\phi$ and 
$\chi_i$.

We now illustrate free-end-point variation in a simple scale-invariant $N$-particle model 
\cite{jb}. The corrected coordinates are $\tilde q_{(i)}=aq_{(i)}$ and the corrected velocities are 
$\dot{\tilde q}_{(i)}=\dot aq_{(i)}+a\dot q_{(i)}$, 
${\mathcal L}={\mathcal L}(\tilde{q}_{(i)},\dot{\tilde q}_{(i)})$. 
Here $a$ is the scaling auxiliary and $q_{(i)}$ are the Cartesian coordinates of unit-mass point 
particles. The single auxiliary $a$ matches the single scaling degeneracy of the $q_{(i)}$. It 
doubles the degeneracy. The habitual pairing of $q_{(i)}$ and $a$ has several consequences, 
including full gauge invariance of all the theory's equations, as we shall show elsewhere. 
More importantly, it imposes constraints. Indeed, the canonical momenta, 
$p^{(i)}=\partial\mathcal L/\partial\dot{q}_i$ and 
$p^{a}=\partial\mathcal L/\partial\dot a$, are
\begin{equation}
\begin{split}
	 p^{(i)} &= \frac{\partial\mathcal L}
			{\partial\dot{\tilde q}_{(i)}}
		\frac{\partial\dot{\tilde q}_{(i)}}
			{\partial\dot{q}_{(i)}} 
		= \frac{\partial\mathcal L}
			{\partial\dot{\tilde q}_{(i)}}a,
	\\p^{a} & = \frac{\partial\mathcal L}
			{\partial\dot{\tilde q}_{(i)}}
		\frac{\partial\dot{\tilde q}_{(i)}}
			{\partial\dot a}
		= \frac{\partial\mathcal L}
			{\partial \dot{\tilde q}_{(i)}}q_{(i)}.
\end{split}
\end{equation}
We therefore have the {\it{primary constraint}}
\beq 
	q_{(i)}p^{(i)}\equiv ap^{a}.
\label{w1}
\eeq
The variation 
$\int\textrm d\lambda(\delta\mathcal L/\delta a)\delta a$ is
\beq
	\int\textrm d\lambda
			\left(\frac{\textrm{d}p^{a}}
					{\textrm{d}\lambda}
 		- \frac{\partial\mathcal L}{\partial a}
			\right)\delta a 
		+ p^{a}(\delta a)_{\textrm{final}} 
		- p^{a}(\delta a)_{\textrm{initial}}.
\label{w2}
\eeq
By the free-end-point rules, this must vanish for any $\delta a$.  
Starting with a general variation 
that vanishes at the end-points, we deduce the standard Euler--Lagrange equation for $a$: 
$\textrm{d}p^{a}/\textrm{d}\lambda=\partial \mathcal L/\partial a$. Then, choosing a variation that does not vanish at the initial end-point but does vanish at the final, 
and enforcing the Euler--Lagrange equations, we deduce that 
$p^{a}_{\textrm{initial}} = 0$.  If we then choose 
a variation that does not vanish at the final end-point, we find that 
$p^{a}_{\textrm{final}} = 0$.  As we can freely 
choose the final point along the curve, we deduce that 
$p^{a}=0$.  The 
Euler-Lagrange equation for $a$ reduces to 
$\partial \mathcal L/\partial a=0$.  This is a key {\it{consistency condition}}. Hence, one can effectively 
minimize the action with respect to $a$ and $\dot a$ independently, although the conditions really 
arise from a free-end-point variation of $a$.  Finally, we get a {\it secondary constraint} 
$q_{(i)}p^{(i)}=0$ from the primary constraint.  These simple considerations capture all the novel 
features of our theory.

We may now proceed to apply this variational procedure to the conformalized BSW action~\eqref{action}.  
Variation with respect to $\dot g_{ij}$ defines the canonical momenta $\pi^{ij}$: 
\beq  
    \pi^{ij} \equiv \frac{\delta I}{\delta\dot{g}_{ij}} 
    = \Pi^{ij} 
	+ \frac{<\tilde\Pi>}{3}\sqrt{g}\,g^{ij}(1 - \wh{\phi}^{6}),
\label{blah} 
\eeq 
where
\beq
\begin{split}
	\Pi^{ij} = \frac{\sqrt{g}}{2N}
    		G^{ijkl}\frac{d \wh{\phi}^{4}g_{kl}}{d\lambda},
	\\N \equiv \frac{1}{2}\sqrt{\frac{\wh{T}}
                    {R - 8\frac{\grad^{2}\phi}{\phi}}},
\label{tarver} 
\end{split}   
\eeq
and $\Pi = \sqrt{g}\,\tilde\Pi=g_{ij}\Pi^{ij}$.  Variation with respect to $\dot\phi$ gives
\beq
    p_{\phi} \equiv \frac{\delta I}{\delta \dot{\phi}} 
        = \frac{4}{\phi}(\Pi - \sqrt{g}\,\wh{\phi}^6<\tilde\Pi>).
\eeq
These definitions imply a primary constraint:
\beq
    p_{\phi} = \frac{4}{\phi}(\pi - \sqrt{g}<p>), 
\label{aumom}
\eeq
where $\pi = \sqrt{g} p = g_{ij}\pi^{ij}$.
Also, the rules of free-end-point variation imply that $p_\phi=0$.  
Hence we find the secondary constraint 
\beq
    p = C, 
\label{CMC}
\eeq 
where $C$ is some spatial (ie $\lambda$-dependent) constant.
This constraint is equivalent to the CMC condition imposed by York, obtained here as a genuine 
constraint and not a gauge-fixing condition (a similar result appeared earlier in~\cite{bom}). 
As this condition designates a preferred time-function, it introduces a notion of global simultaneity. 

The definitions of the momenta imply a quadratic primary identity, the analogue of the GR Hamiltonian 
constraint:
\begin{multline} 
    \pi^{ij}\pi_{ij} -\frac{\pi^2}{2} 
	- \frac{C^2}{6}(1 - \wh{\phi}^{6})^{2} 
	+ \frac{C}{3}\sqrt{g}\pi(1 - \wh{\phi}^{6}) 
	\\ = g\wh{\phi}^{8}
		\biggl(R - 8\frac{\grad^{2}\phi}{\phi}\biggr).
\label{ham1}
\end{multline}
Defining 
$\sigma^{ij} \equiv \pi^{ij} - \frac{1}{3}g^{ij}\pi$, and enforcing the CMC constraint~\eqref{CMC}, eqn.~\eqref{ham1} takes the form 
\beq
    \sigma^{ij}\sigma_{ij} - \frac{\pi^{2}\wh{\phi}^{12}}{6} 
    	- g\wh{\phi}^{8}
		\biggl(R - 8\frac{\grad^{2}\wh\phi}{\wh\phi}\biggr) 	= 0.
\eeq
This identity holds true on any path in the configuration space.  If we, however, switch to a 
Hamiltonian picture in which the canonical momenta become independent variables, it becomes, just as 
in standard GR, a real constraint.  It is then clearly the Lichnerowicz--York equation~\cite{yor}.  

The variation wrt $\xi_{i}$ gives
\beq
    \grad_{j}\pi^{ij} = 0,\label{mom} 
\eeq
where we have imposed the CMC constraint~\eqref{CMC}; this relation is formally identical to the standard GR momentum 
constraint. 

Variation with respect to $\phi$ and imposition of~\eqref{CMC} 
leads to the 
consistency condition
\beq 
    \wh{\phi}^4 N \left(R - 7\frac{\grad^{2}\phi}{\phi}\right) 
        - \wh\phi\grad^{2}\left(\wh{\phi}^3N \right) 
        + \wh{\phi}^8\frac{N p^{2}}{4} = \wh{\phi}^6 D,
\label{LFE}  
\eeq 
where
\beq
    D = \biggl<\wh{\phi}^{4}N
		\biggl(R - 8\frac{\grad^{2}\phi}{\phi}\biggr) 
        + \frac{\wh{\phi}^8 N p^{2}}{4}\biggr>.
\eeq
One can show that $D$ is the global average of the 
left-hand side of eqn.~\eqref{LFE}.  

We now make a dimensional analysis. We take 
the metric, and by extension $\phi$, to be 
dimensionless and give dimensions to the coordinates.
We set the speed of light equal to one, 
so all coordinates have dimensions
of length (\textit{l}).  Counting derivatives 
in their definitions, we find
the dimensions of other objects: 
$[R] = [\wh R] = [\wh T] = 1/l^2$.  
The lapse, $N$, is dimensionless.  It follows that 
          $[\pi^{ij}] = [\pi] = [p] = [C] = 1/l$.  In the York scheme,
one fixes an implicit length scale 
for the CWB manifold by specifying a numerical
value for the dimensionful parameter $C$.  That the scheme
determines a physical 3-metric from an initially 
unphysical one reflects this fixing, as the manifold
acquires a volume corresponding to the implicit length scale.

Our approach is complementary to this. We specify as initial data $g_{ij}, \dot{g}_{ij}$. Our procedure subjects them to considerable `gauge dressing', but the initial volume $V$ corresponding to the initial $g_{ij}$ is the one thing that must not be changed. So we put the scale in explicitly through the canonical coordinate $g_{ij}$, not though a combination of the canonical momenta. In both cases, the specification of the numerical value of a single quantity with a length dimension fixes the numerical values of all other quantities with a length dimension. The time label is fixed only up to global reparametrization (as~\eqref{LFE} is homogeneous in $N$), and, as always, the spatial coordinates are freely specifiable.

Our calculations have given us five equations---the CMC condition~\eqref{CMC}, the momentum constraint~\eqref{mom}, 
and the consistency condition~\eqref{LFE}---that we try to solve for the five gauge variables 
$(\wh\phi, \partial\wh\phi/\partial\lambda, \xi_i)$ given the initial $g_{ij}, \dot{g}_{ij}$.  In accordance with our dimensional analysis the constant $C$, with its dimension $l^{-1}$, should emerge as an 
eigenvalue for this problem.  There are many choices of $(g_{ij}, \dot g_{ij})$ for which we know 
solutions exist; e.g.~take a CMC slice through a space-time that solves the Einstein equations, 
take the physical $g_{ij}$ and $\dot g_{ij}$ given by the CMC foliation, and multiply by any VPCT.  
On the other hand, with a given arbitrary $(g_{ij}, \dot g_{ij})$, we have no existence theorem for 
solutions.  This problem is very similar to the thin-sandwich problem in GR~\cite{bsw, ts}. 

If and when we can solve the equations, we can switch to the `corrected coordinates' 
$\overline g_{ij}$.  In this frame, $\wh\phi = 1$, and the consistency condition~\eqref{LFE} reduces 
to
\beq
    N R - \nabla^2 N + \frac{N p^2}{4} = D.
\eeq
This relation is the CMC lapse-fixing equation of GR; satisfying it guarantees the propagation of the 
CMC constraint~\eqref{CMC}.

In contrast to the point of view of this article, York chose to retain 
the spacetime ontology and set up his powerful conformally-invariant 
method using invariance of the decoupling of the ADM Hamiltonian and momentum constraints 
as guiding principle. This led him to regard the Lichnerowicz--York equation as a gauge fixing 
(see especially his comments near the end of the first paper in~\cite{yor}).  From our 
point of view, we rather choose to interpret the initial value problem as containing a 
genuine gauge invariance, which leads to gauge corrections both to the potential and to  
the metric velocities.  From our point of view, we have derived GR, 
a prescription for solving its initial-value problem, and the condition for maintaining the 
CMC condition, all in a single package from our first principles. The transition to 
the Hamiltonian formulation and the implications of our results will be considered elsewhere.
   
\begin{acknowledgments}
We dedicate this paper to Jimmy York on the occasion of his 65th birthday. 
The authors thank him for discussions. 
EA also thanks Malcolm MacCallum for discussions and PPARC for financial support. 
EA, BF, BK, and N\'OM thank the Barbour family for hospitality.
\end{acknowledgments}

\end{document}